# Configurational entropy of magnetic skyrmions as an ideal gas

R. Zivieri[1*], R. Tomasello[2#], O. Chubykalo-Fesenko[3], V. Tiberkevich[4], M. Carpentieri[5], and G. Finocchio[1§]

[1]*Department of Mathematical and Computer Sciences, Physical Sciences and Earth Sciences, University of Messina, 98166 Messina, Italy*

[2]*Institute of Applied and Computational Mathematics, Foundation for Research and Technology, GR 700 13 Heraklion, Crete, Greece*

[3]*Instituto de Ciencia de Materiales de Madrid, CSIC, Cantoblanco, Madrid, Spain*

[4]*Department of Physics, Oakland University, Rochester, Michigan 48309, USA*

[5]*Department of Electrical and Information Engineering, Politecnico di Bari, Bari, Italy*

*zivieri@fe.infn.it

#rtomasello@iacm.forth.gr

§gfinocchio@unime.it

## Abstract

The study of thermodynamics of topological defects is an important challenge to understand their underlying physics. Among them, magnetic skyrmions have a leading role for their physical properties and potential applications in storage and neuromorphic computing. In this paper, the thermodynamic statistics of magnetic skyrmions is derived. It is shown that the skyrmion free energy can be modelled via a parabolic function and the diameters statistics obeys the Maxwell-Boltzmann distribution. This allows for making an analogy between the behavior of the distribution of skyrmion diameters statistics and the diluted gas Maxwell-Boltzmann molecules distribution at thermodynamical equilibrium. The calculation of the skyrmion configurational entropy, due to thermally-induced changes of size and shape of the skyrmion, is essential for the determination of thermal fluctuations of the skyrmion energy around its average value. These results can be employed to advance the field of skyrmionics.

# I. INTRODUCTION

Magnetic skyrmions have been gaining an important role in studies of low-dimensional magnetic systems due to their suitable physical properties and potential applications [1–4]. Skyrmions can be considered as quasi-particles with topologically-protected magnetization texture, characterized by an integer skyrmion number [1,4]. Although skyrmions can be stabilized by the interplay between exchange and dipolar interactions (so called "bubble skyrmion" [1]), much of the interest is devoted to systems where the Dzyaloshinskii–Moriya interaction (DMI) plays a role in this stabilization [5,6]. The DMI is a chiral exchange interaction due to lack or breaking of inversion symmetry in bulk crystalline lattices (bulk DMI) [7–9] or at the interfaces in magnetic multilayers (interfacial DMI (IDMI)) [10–13]. While other types of DMI can exist (for instance, in $D_2d$ structures) [14,15], in this work we focus on the IDMI. This because it promotes the formation of small Néel skyrmions [1,2,4], which are stable at room temperature as isolated skyrmions, and can be nucleated [16–18], manipulated [12,19–21] as well as detected [22–24] by electrical currents. Therefore, Néel skyrmions have become promising for technological applications [25–33]. Fundamentally, because of thermal fluctuations, Néel skyrmions are subject to (i) internal deformations [10,12,34,35], that are responsible for the loss of the circular symmetry; (ii) thermal drift [18,34,35], which leads to a random skyrmion motion throughout the film plane; (iii) thermal breathing modes [35,36] that can induce non-stationary expansion and shrinking of the skyrmion core, i.e. a time-evolution of the skyrmion size. Hence, the effect of thermal fluctuations should be considered for a proper design of skyrmion-based devices and applications [32,35], especially at room temperature.

In this work, we show that the thermal fluctuations promote the existence of a number of skyrmions characterized by the same energy, but having different shapes and diameters. This aspect allows us for the definition of a skyrmion configurational entropy by using a statistical thermodynamic analogy between the skyrmion diameter population and the non-interacting molecules of an ideal gas [37,38]. This approach is based on the analytical formulation previously developed [35] considering region of parameters where the two following hypotheses are verified: (i) the skyrmion energy can be well approximated by a square function of the skyrmion diameter near the minimum, and (ii) the skyrmion diameter distribution is well-described by a Maxwell-Boltzmann (MB) function. The validity of those two hypotheses is checked by taking advantage of full micromagnetic simulations for different combinations of temperature and external field. The skyrmion average diameter and its standard deviation, as well as the skyrmion entropy, can be analytically derived. In addition, the developed model can also be extended to the description of further magnetic textures, such as bubbles and vortices.

From a theoretical point of view, the knowledge of the skyrmion configurational entropy allows us to construct the proper thermodynamics for determining the fluctuations for many important quantities, such as the free energy. We show that a distribution of skyrmion diameters can be described in a canonical ensemble characterized by fluctuations of energy that are much smaller than the skyrmion average energy.

Up to date, the skyrmion entropy has been estimated from the experimental data, and it has been used to characterize the type (first- or second-order) of magnetization phase-transition around the transition temperature in bulk B20 compounds [39,40], and as a corrective term to the Arrhenius law, to explain the discrepancy between experimental and theoretical calculations of the lifetime of a skyrmion lattice [41]. However, no direct dependence of the entropy on the physical parameters, temperature and geometrical characteristics of the skyrmion has been expressed. This work will fill this gap giving a simple analytical model to be used as a support for the experimental works at finite temperatures.

## II. NUMERICAL MODEL

We start by performing micromagnetic simulations at finite temperature to generate the data to be analyzed. We consider a circular nanodot of diameter $2R_d$ =400 nm of a ferromagnetic material (we consider Cobalt here) with a thickness of 0.8 nm assumed to be coupled with a thin layer of heavy metal giving a sufficiently-large IDMI, i.e. Platinum. We perform systematic micromagnetic simulations to calculate the skyrmion sizes as a function of the out-of-plane external field $\mu_0 H_{ext}$ and temperature $T$, by integrating the Landau-Lifshitz-Gilbert equation for the reduced magnetization $\mathbf{m} = \mathbf{M}/M_s$ [42–48] ($M_s$ is the saturation magnetization, see note 1 in the Supplemental Material). At $T$=0 K, we used the following material parameters: $M_s$=600 kA/m [12], $A$=20 pJ/m [49], $D$=3.0 mJ/m$^2$ [50,51], $K_u$=0.60 MJ/m$^3$ [12,52], and Gilbert damping $\alpha_G$=0.1 [53], while for $T$>0 and the analytical model, we used the parameters as calculated from the scaling relations [35] $A(m) = A(0)m^{1.5}$, $D(T) = D(0)m(T)^{1.5}$. and $K_u(T) = K_u(0)m(T)^{3.6}$.

For the micromagnetic simulations, the thermal effects are included in the LLG equation as a stochastic term $\mathbf{h}_{th}$ added to the deterministic effective magnetic field in each computational cell $\mathbf{h}_{th} = (\chi/M_S)\sqrt{2(\alpha K_B T / \mu_0 \gamma_0 \Delta V M_s \Delta t)}$, with $K_B$ being the Boltzmann constant, $T$ temperature of the sample, $\mu_0$ the vacuum permeability, $\gamma_0$ the gyromagnetic ratio, $\Delta V$ the volume of the computational cell, $\Delta t$ the simulation time step, and $\chi$ a three-dimensional white Gaussian noise with zero mean and unit variance [47,48]. The thermal fields, in each computational cell, are uncorrelated. The discretization cell size used is 2.5x2.5x0.8 nm$^3$ [42] (see note 1 in the Supplemental Material for more details).

The effective diameter is calculated by assuming that the area of the skyrmion core (here it is the region where the *z*-component of the magnetization is negative) is equivalent to a circle [54].

Figure 1(a) shows the time dependence of the total micromagnetic energy $E[\mathbf{m}] = \int dV \varepsilon(\mathbf{m})$ of the ferromagnet ($\varepsilon(\mathbf{m})$ is the energy density) as calculated by micromagnetic simulations from the spatial distribution of the magnetization $\mathbf{m}$:

$$\varepsilon(\mathbf{m}) = A(\nabla \mathbf{m})^2 + D\left[m_z(\nabla \cdot \mathbf{m}) - (\mathbf{m} \cdot \nabla)m_z\right] + K_u\left(1 - m_z^2\right) - 0.5 M_s \mathbf{m} \cdot \mathbf{H}_m - M_s \mathbf{m} \cdot \mathbf{H}_{ext}, \quad (1)$$

where $m_z$ is the magnetization *z*-component, $\mathbf{H}_m$ is the magnetostatic field, and $\mathbf{H}_{ext}$ is the external magnetic field.

It can be observed that there exist skyrmion configurations with different shape and size (Figs. 1(b)-(e)), but characterized by the same energy (5.6 x 10$^{-17}$ J in Fig. 1(a)). These results suggest that a skyrmion configurational entropy can be introduced as the number of different skyrmions having the same energy.

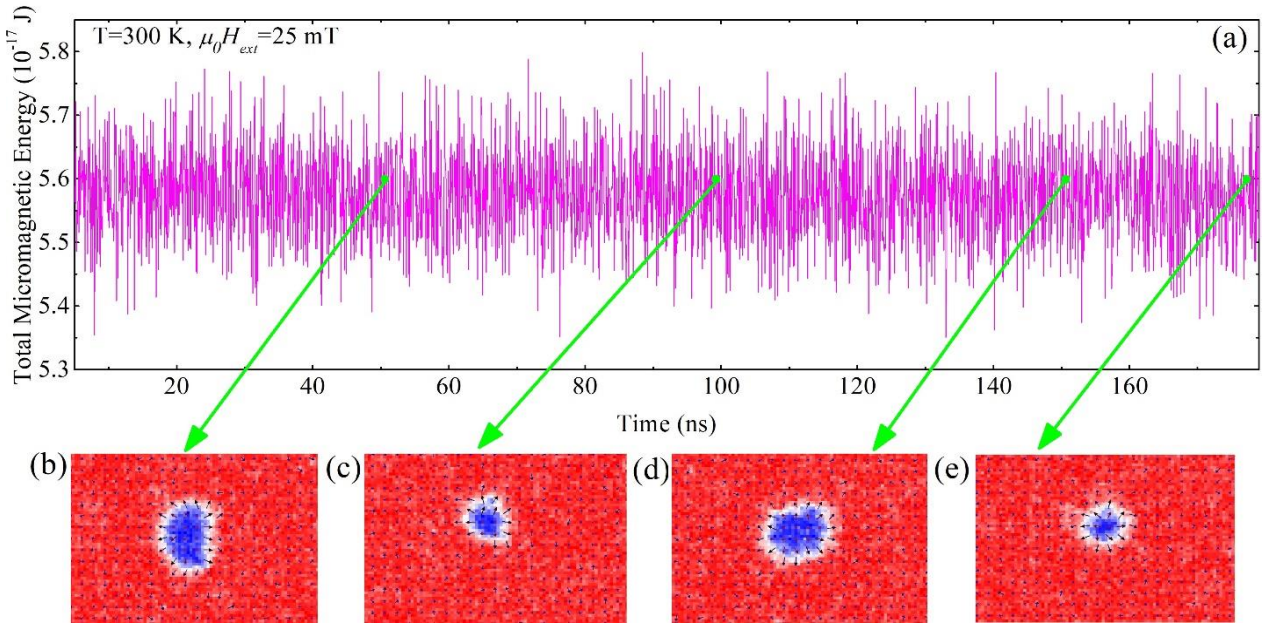

Fig. 1: (a) Total micromagnetic energy as a function of time for one of the simulations (*T*=300 K, and $\mu_0 H_{ext}$=25 mT). The four green circles indicate an example of time instants at which different skyrmions have the same energy. (b)-(e) Snapshots of the skyrmion at the time instants indicated in (a).

## III. ANALYTICAL MODEL

### A. Skyrmion energy

The computation of the configurational entropy is based on the determination of the skyrmion energy. The skyrmion magnetization texture in cylindrical coordinates can be written in the form $\mathbf{m}(x,y) = \sin\theta(\rho)\cos\phi_0 \hat{\rho} + \sin\theta(\rho)\sin\phi_0 \hat{\phi} + \cos\theta(\rho)\hat{z}$ where $\theta(\rho)$ is the radially dependent magnetization distribution angle and $\phi_0$ is an azimuthal angle. Setting $\phi_0 = 0$, we get the outwardly directed radial Néel skyrmion texture, $\mathbf{m}(x,y) = \sin\theta(\rho)\hat{\rho} + \cos\theta(\rho)\hat{z}$. In the present study, we have considered an outwardly radial Néel skyrmion with a negative core, that is characterized by a skyrmion number $Q = -1$, a cylindrical symmetry with respect to the out-of-plane direction (z-axis), and it is stabilized in a magnetic circular dot. The results derived for Néel skyrmion can indeed be generalized to other magnetic textures.

The skyrmion energy $E$ is calculated as a volume integral of the skyrmion energy density $E(r_{sk}) = \int \varepsilon(r, r_{sk}) dV = \int_{-t/2}^{t/2} dz \int_0^{2\pi} d\phi \int_0^{R_d} \varepsilon(r, r_{sk}) \rho d\rho$. For the Néel skyrmion, we use the following trial equilibrium magnetization distribution through the ansatz $\theta_0(r) = 2\arctan[\frac{r_{sk}}{r} e^{\xi(r_{sk}-r)}]$, with $r_{sk} = R_{sk}/l_{exch}$ the dimensionless skyrmion radius, $l_{exch} = \sqrt{2A/\mu_0 M_s^2}$ is the exchange length, with $A$ the material exchange stiffness, $t$ the dot thickness and $\rho = r/l_{exch}$ [35,55]. This skyrmion ansatz has been previously shown to have an excellent agreement with direct micromagnetic simulations [35,56]. The skyrmion energy density $\varepsilon = \varepsilon_{exch} + \varepsilon_{IDMI} + \varepsilon_{ani} + \varepsilon_{ext}$ contains all the relevant contributions. In particular, $\varepsilon_{exch} = A(\nabla \mathbf{m})^2$ is the exchange energy density, $\varepsilon_{IDMI} = |D|[m_z(\nabla \cdot \mathbf{m}) - (\mathbf{m} \cdot \nabla)m_z]$ is IDMI energy density and $D$ the DMI strength, $\varepsilon_{ani} = K_u(1 - m_z^2) + \frac{1}{2}\mu_0 M_s^2 m_z^2$ is the anisotropy energy density with $K_u$ the uniaxial anisotropy constant, and $\varepsilon_{ext} = -M_s B m_z$ is the Zeeman energy density, with $B = \mu_0 H_{ext}$ being the amplitude of the external bias $\mathbf{B}$ parallel to the z-axis. The magnetic parameters $A$, $D$ and $K_u$ at non-zero temperature are scaled from their zero temperature values, by using the scaling laws shown in the Section II.

The first hypothesis of our analytical approach is that the skyrmion energy near the minimum can be described via a parabolic potential of the form $E_{sky} \simeq a(D_{sky} - D_{0\,sky})^2 + b$. In the parabolic potential, the coefficient $a$ [J/m$^2$] is proportional to the parabola curvature, $a = \frac{1}{2}\, d^{\,2}E_{sky}/dD_{sky}^{\,2}$, $D_{sky}$ is the generic skyrmion diameter ranging from 0 to $2R_d$, $D_{0\,sky}$ is the equilibrium diameter corresponding to the energy minimum $E_{sky}^{min}$ for every $T$, and $b$ [J] is $b = E_{sky}^{min}$. The equilibrium skyrmion diameter depends on the temperature via the scaled values of the magnetic parameters, therefore, both $a$ and $b$ are temperature-dependent coefficients, $a(T)$ and $b(T)$. Figure 2(a) shows, in fact, how the coefficient $a$ changes with temperature, for three values of the external field. The general trend is a linear dependence of $a$ on the temperature and $a$ decreases with increasing temperature marking a broadening of the potential well due to thermal effects. Moreover, at fixed temperature, $a$ increases with increasing the external field amplitude indicating its narrowing effect on the potential well.

Figure 2(b) and (c) show that the parabolic curve fits well the analytical skyrmion energy when $0 \leq T \leq 200$ K at zero external field, while, at $T=300$ K (Fig. 2(d)), the matching is less accurate. We ascribe this difference to the change of the energetic stability of the skyrmion when it is becoming the ground state (see Fig. 3 in Ref. [35]). In particular, for the parameters used in this study, the skyrmion is a metastable state when either $0 \leq T \leq 200$ K for any external field, or $0 \leq T \leq 300$ K for $\mu_0 H_{ext} > 5$ mT, while it becomes stable outside these intervals. As was explained in Ref. [35] with details, during this transition the skyrmion radius is very sensitive to small variations of external parameters.

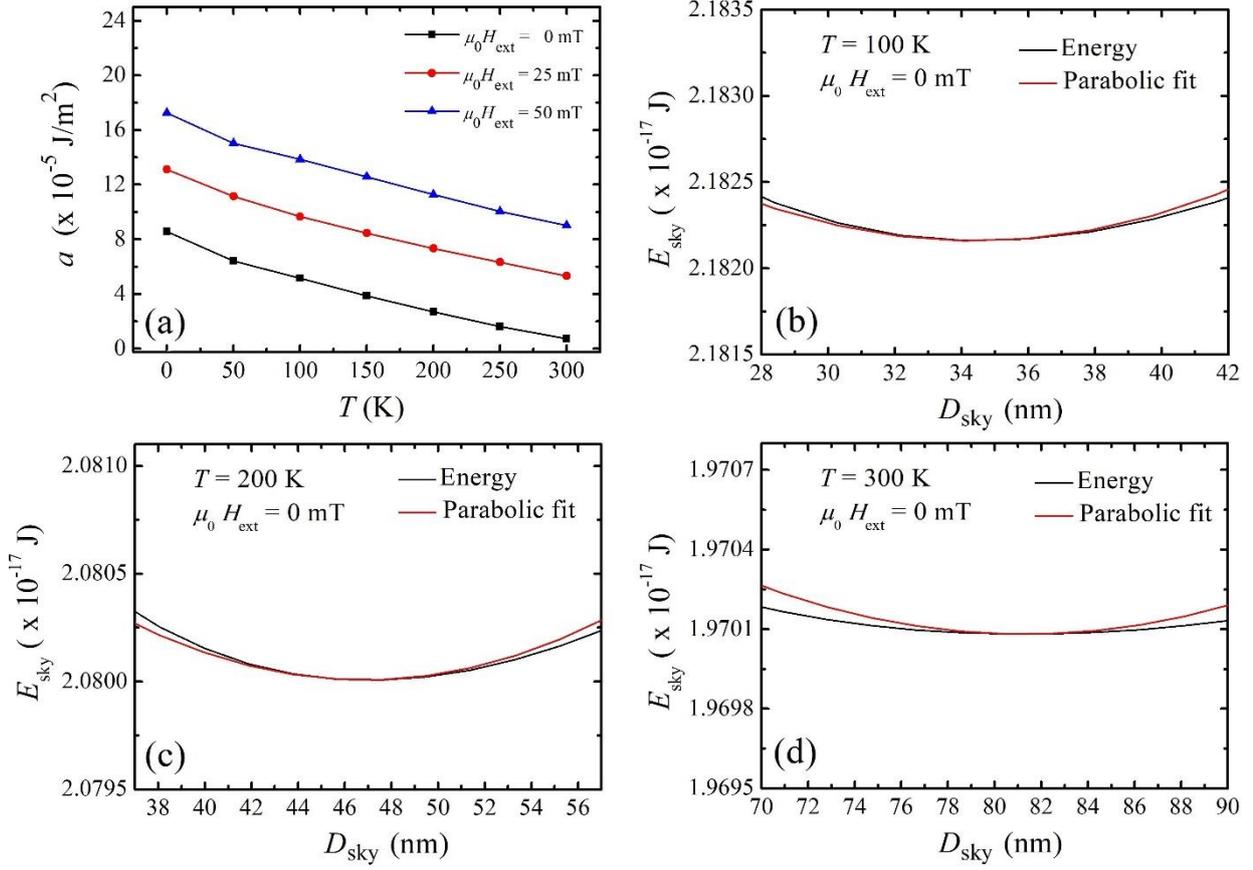

Fig. 2: (a) Fitting parameter *a* as a function of $T$ for $\mu_0 H_{ext}$= 0 mT, 25 mT, 50 mT. Profile of the skyrmion energy close to the energy minimum (black curve) together with the parabolic fit (red curve) for $\mu_0 H_{ext}$= 0 mT and (b) $T$ = 100 K, (c) $T$=200 K, and (d) $T$=300 K.

### B. Skyrmion diameter distribution

The fact that the energy of the skyrmion is, as a first approximation, well-described by as a quadratic function of the skyrmion diameter leads us to suppose that the skyrmion diameter distribution can be treated as the particles of an ideal gas, at least from a statistical thermodynamics viewpoint. Therefore, the second step of our analytical approach is to check if the population of the skyrmion diameters follows a Maxwell-Boltzmann distribution function:

$$\frac{dn}{dD_{sky}} = C_{sky} \, D_{sky}^2 \, e^{-\beta a \left(D_{sky} - D_{0\,sky}\right)^2}, \qquad (2)$$

where $C_{sky}$ is the normalization constant, and $\beta = \dfrac{1}{k_B T}$ with $k_B$ = 1.38 × 10$^{-23}$ J/K the Boltzmann constant [42] (see note 2 in the Supplemental Material). Analogously to the ideal gas, the skyrmion energy is comparable with the thermal energy in the range of temperatures 50 ÷ 300 K, thus ensuring a displacement effect of the maximum of the distribution [37].

In order to prove this hypothesis, we compare the distribution of skyrmion diameter as computed from micromagnetic simulations with Eq. (2). Figure 3 shows such a comparison at $T = 100, 200$ and 300 K, respectively, for an applied field $\mu_0 H_{ext} = 25$ mT. The agreement between the analytical and the micromagnetic results is excellent. Similar good agreements are also obtained for the other temperature/external field combinations in the region of metastability.

Since we have shown that the skyrmion energy can be considered as a quadratic function of the skyrmion diameter and that the skyrmion diameter population obeys to the MB distribution, we can make the analogy between the behavior of skyrmion diameter population and the one of non-interacting molecules of an ideal gas. Let us examine in depth the point later in the text.

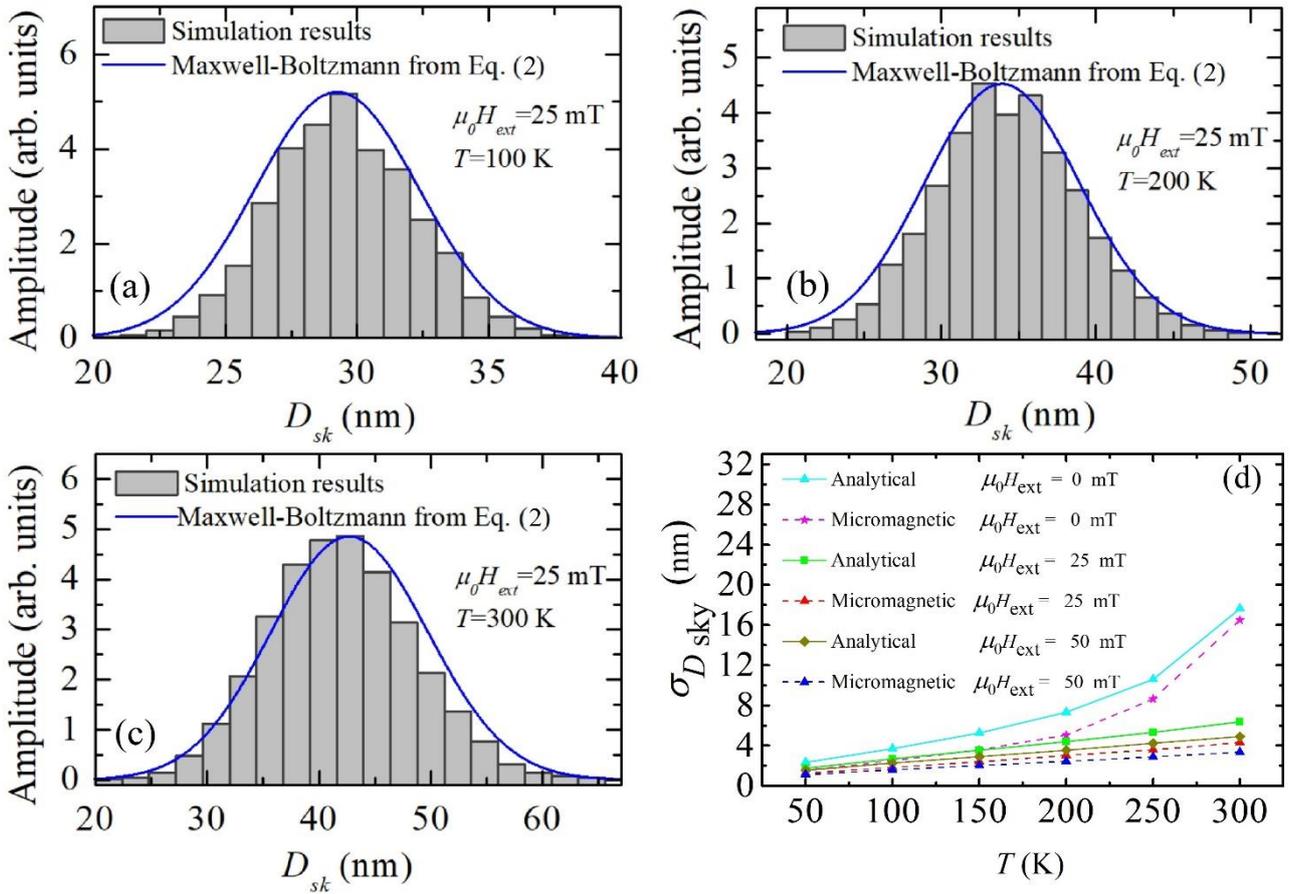

Fig. 3: Skyrmion diameter distribution for (a) $T = 100$ K, (b) $T = 200$ K, and (c) $T = 300$ K when $\mu_0 H = 25$ mT. The histograms represent the results from micromagnetic simulations with thermal fields, while the blue curve is the analytical MB distribution as calculated from Eq. (2). In the analytical calculations, we used the scaled values of the magnetic parameters $A$, $D$, and $K_u$ for each temperature [35] and obtained the following parameters: at $T = 100$ K, $a = 9.67 \times 10^{-5}$ J/m$^2$ and $D_{0\ sky}$ = 26.83 nm, at $T = 200$ K, $a = 7.33 \times 10^{-5}$ J/m$^2$ and $D_{0\ sky}$ = 32.03 nm and at $T = 300$ K, $a = 5.32 \times 10^{-5}$ J/m$^2$ and $D_{0\ sky}$ = 39.64 nm. (d) Standard deviation as a function of temperature, for three values of the external field. The continuous lines with symbols represent the analytical calculations by means of Eq. (3), while the dashed lines with symbols indicate the results from micromagnetic simulations with thermal stochastic fields.

In Table 1, we strengthen this analogy considering a skyrmion diameter statistics and an ideal gas of $N$ non-interacting molecules. Straightforwardly, the skyrmion diameter $D_{sky}$ replaces the velocity v of the gas molecule, while the fitting parameter $a$, proportional to parabolic energy curvature, plays the same role as the constant ½ $m$ depending on the mass $m$ of the molecule in the gas. In this way, the MB distribution reproducing the statistical behavior of skyrmion diameters maps into the well-known MB distribution of velocities in a diluted gas [37].

Table 1. Analogy between an ideal gas of molecules and a magnetic skyrmion diameter population with $C_g = 4\pi N (m/(2\pi k_B T))^{3/2}$.

| Ideal gas | Skyrmion diameters |
|---|---|
| v | $D_{sky}$ |
| ½ $m$ | $a$ |
| $\frac{dn}{dv} = C_g v^2 \exp(-1/2\ m\ v^2/k_B T)$ | $\frac{dn}{dD_{sky}} = C_{sky} D_{sky}^2 \exp(-a (D_{sky} - D_{0\ sky})^2 / k_B T)$ |

### C. Average skyrmion diameter

Thanks to the aforementioned analogy, we can calculate the average skyrmion diameter

$$<D_{sky}(T)> = \frac{\int_0^\infty dD_{sky} D_{sky}^3 e^{-\beta a (D_{sky} - D_{0\ sky})^2}}{\int_0^\infty dD_{sky} D_{sky}^2 e^{-\beta a (D_{sky} - D_{0\ sky})^2}}$$ at a given temperature, in the same way as the average

particle velocity in an ideal dilute gas:

$$<D_{\text{sky}}(T)> \simeq D_{0\,\text{sky}} \frac{3k_B T + 2a D_{0\,\text{sky}}^2}{k_B T + 2a D_{0\,\text{sky}}^2}. \tag{3}$$

### D. Standard deviation of the skyrmion diameter distribution

The skyrmion average diameter is crucial to compute: 1) the standard deviation $\sigma_{<D_{\text{sky}}>}$ of the diameter distribution and 2) the skyrmion configurational entropy $S$. First, we calculate

$$\sigma_{<D_{\text{sky}}>} = \sqrt{\frac{\int_0^\infty dD_{\text{sky}} \left(D_{\text{sky}} - <D_{\text{sky}}>\right)^2 D_{\text{sky}}^2 \, e^{-\beta a \left(D_{\text{sky}} - <D_{\text{sky}}>\right)^2}}{\int_0^\infty dD_{\text{sky}} D_{\text{sky}}^2 \, e^{-\beta a \left(D_{\text{sky}} - <D_{\text{sky}}>\right)^2}}}$$

with $<D_{\text{sky}}> = <D_{\text{sky}}(T)>$ as an integration over the MB distribution, obtaining:

$$\sigma_{<D_{\text{sky}}>}(T) \simeq \sqrt{\frac{k_B T}{2a} \frac{3k_B T + 2a <D_{\text{sky}}>^2}{k_B T + 2a <D_{\text{sky}}>^2}} \tag{4}$$

In Fig. 3(d), we compare the standard deviation as obtained from micromagnetic simulations and Eq. (4), observing that they match well in the region of energetic metastability.

### E. Configurational entropy

We now outline the computation of the skyrmion configurational entropy that represents one of the key results of this study. We wish to remind that the source of this entropy is mainly due to the skyrmion internal deformations and thermal breathing mode [35,36]. We employ the definition of the Boltzmann order function $H_0$ for a dilute ideal gas [38], that represents a measure of order and it is proportional to the MB distribution, that is the solution of the Boltzmann equation at thermodynamic equilibrium. Indeed, $H_0$ is a quantity defined as the opposite of the entropy $S$ at equilibrium, namely $H_0 = -S/k_B$ [38], with $H_0 < 0$. $S$ gives the direct connection between the canonical ensemble and thermodynamics. In the continuous case applied to our framework, where minor changes of the skyrmion size occur along the radial direction, $H_0$ can be written as a functional integral representing the statistical average $< \ln f_0 >$ over all spatial directions:

$$H_0 = \frac{\pi}{2} \int_0^\infty dD_{\text{sky}} D_{\text{sky}}^2 \, f_0 \ln f_0, \tag{5}$$

with $f_0 = C_{sky} e^{-\beta\left[a\left(D_{sky}-<D_{sky}>\right)^2\right]}$ being the Gaussian distribution of the skyrmion diameters at thermodynamic equilibrium, that has the meaning of a probability density in statistical mechanics. The skyrmion configurational entropy turns out to be ($S = -k_B H_0$):

$$S \simeq k_B \left[ \ln\left(\frac{(k_B T)^{\frac{3}{2}} + 2(k_B T)^{\frac{1}{2}} a <D_{sky}>^2}{a^{\frac{3}{2}} <D_{sky}>^2 t}\right) + \frac{1}{4}\left(\frac{3k_B T + 2a <D_{sky}>^2}{k_B T + 2a <D_{sky}>^2}\right)\right] + S_0, \qquad (6)$$

with $S_0 = \frac{1}{2} k_B \ln \pi$ a constant (see Appendix A for the details of the calculations leading to Eq. (6)).

As expected, the configurational entropy has a geometric, thermal, and magnetic parameters dependence. $S$ depends on the size of the skyrmion via $<D_{sky}>$, confirming the link with the thermal breathing mode (Fig. 1(b)-(e)), decreases with decreasing temperature, denoting the strict connection with temperature effects until a minimum temperature close to 1K (see the next section for more details), as well as depending on the magnetic parameters via the coefficient $a$. Note that Eq. (6) is a general result since it is independent of the chosen skyrmion distribution texture. Indeed, different skyrmion distribution texture would lead to energy profiles again reproducible in the neighborhood of the minimum by means of a parabolic dependence.

However, as $T \to 0$, we get, from Eq. (6), $S \to -\infty$ apparently contradicting Nernst's theorem or third principle of thermodynamics according to which entropy of a crystalline body equals zero at absolute zero temperature. This result is not surprising and agrees with the well-known one of the Sackur-Tetrode entropy equation for an ideal gas [37]. Indeed, in both cases the derivation is classical, resulting from the application of the classical MB statistics.

Figure 4(a) shows the skyrmion configurational entropy, calculated according to Eq. (6), as a function of temperature for three different external fields (0, 25 and 50 mT). The entropy increases with increasing temperature and its increase is more remarkable in the absence of an external bias field. This behavior reflects the higher disorder due to the larger deformations and thermal breathing mode of the skyrmion at room temperature [35]. This disorder is partially reduced by the ordering effect of the external bias field.

Figure 4(b) illustrates the trend of the configurational entropy as a function of the external field at fixed temperature ($T = 300$ K). It is evident the entropy decreases due to the external bias field that leads to a reduction of the disorder of the system.

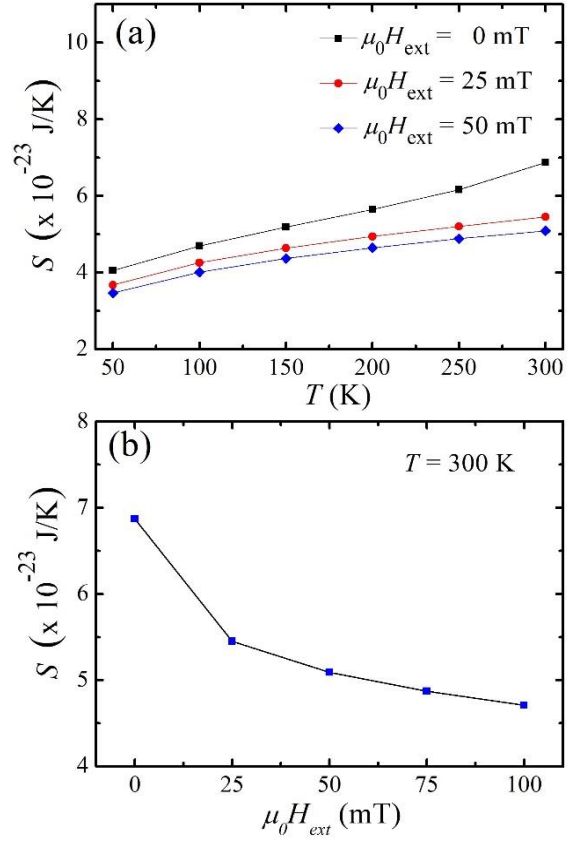

FIG. 4. Configurational entropy as a function of (a) temperature for $\mu_0 H_{ext} = 0$ mT, 25 mT and 50 mT, as indicated in the legend, and (b) the external field at $T = 300$ K.

### F. Behavior of $<D_{sky}>$, $\sigma_{Dsky}$, and $S$ at low temperature

It is interesting to derive the low-temperature behavior of the average skyrmion diameter. We assume, for the whole range of temperatures studied (0 ÷ 300 K) $a(T) = a_0 - c\,T$ (see Fig. 2(d)) with $a_0 = a\,(T = 0\text{ K})$ and $c$ a coefficient expressed in J/m² K, and, at low temperatures, $D_{0\,sky}(T) = D_{0\,sky}(T = 0\text{ K}) + d\,T$, with $d$ a coefficient expressed in m/K.

For $T \to 0$ K, we get, from a numerical calculation, that $<D_{sky}(T \to 0\text{ K})> = D_{0\,sky}(T = 0\text{ K})$. At low temperature, the expansion of Eq. (3) to first order, via $<D_{sky}(T \to 0\text{ K})> = D_{0\,sky}(T = 0\text{ K})$, yields:

$$<D_{sky}(T)>_{T \to 0} \approx D_{0\,sky}(T = 0\,K) + \left(\frac{k_B}{a_0\,D_{0\,sky}(T = 0\,K)} + d\right)T, \qquad (7)$$

with $a_0 = a(T = 0\,K)$.

Eq. (7) expresses, in the regime of low temperatures, a linear dependence of the average skyrmion diameter on temperature, as it can be observed in Fig. 5(a) for different combinations of low temperature and external field.

At low temperatures ($T \to 0$), via a series expansion to the first order and taking into account the abovementioned assumptions on $a$ and $D_{0\,sky}$, the standard deviation turns out to be:

$$\sigma_{<D_{sky}>(T \to 0)} \approx \sqrt{\frac{k_B T}{2a_0}} \qquad (8)$$

Hence, at low temperatures the standard deviation has a square root proportionality on temperature, as shown in Fig. 5(b).

In the same way, the configurational entropy for, $T \to 0$, reads

$$S_{(T \to 0)} \approx k_B \left[ \ln\left( \frac{2}{t}\left(\frac{k_B T}{a}\right)^{\frac{1}{2}} \right) + \left( \frac{3}{4}\frac{k_B}{a_0 <D_{sky}>^2} + \frac{c}{2a_0} \right) T \right] + S_1 \qquad (9)$$

with $S_1 = k_B \left( \frac{1}{2}\ln \pi + \frac{1}{4} \right)$.

The first term on the second member has a logarithmic dependence on $T$, hence giving the divergence of $S$ for $T = 0$ K. On the other hand, the second term expresses the linear dependence of $S$ on $T$ resulting from the expansion of both terms.

Figure 5(c) shows $S$ vs. $T$ for $1 < T < 50$ K. It is evident the deviation from the linear behavior is due to the logarithmic dependence on $T$ of the first term. This trend is similar also in the presence of an external field and is also present at higher temperatures as shown in Fig. 4(a).

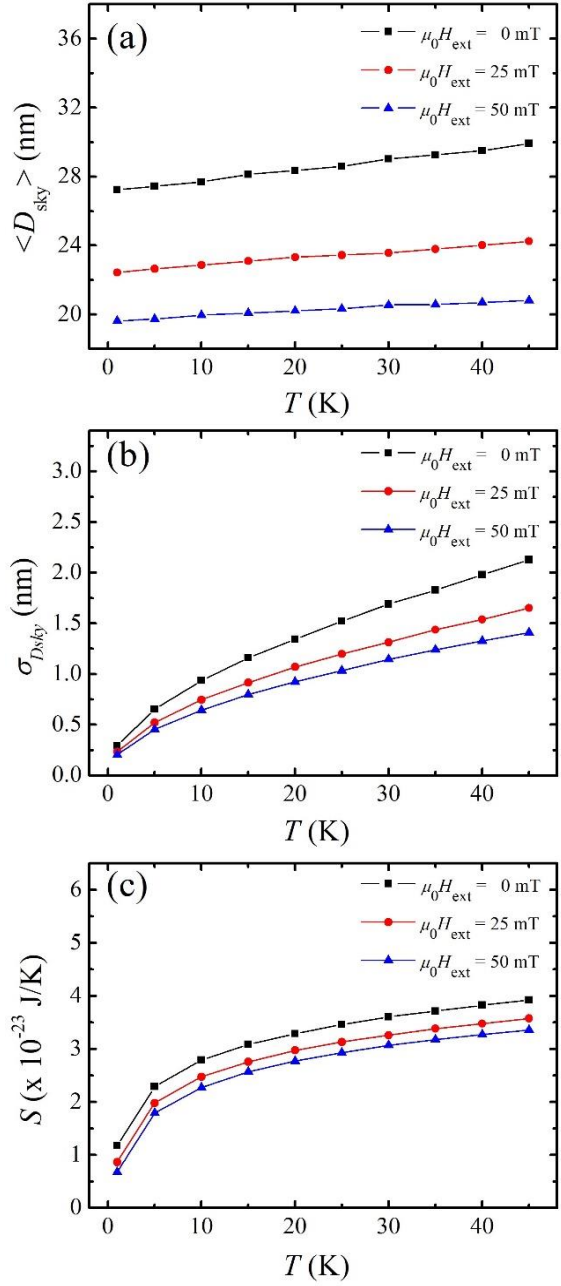

FIG. 5. Low-temperature behavior of (a) average skyrmion diameter, as obtained by Eq. (7), (b) standard deviation of the skyrmion diameter distribution, as obtained by Eq. (8), and (c) configurational entropy, as obtained by Eq. (9).

## G. Energy Fluctuations

The skyrmion configurational entropy is crucial to determine: 1) the magnetization distribution as a function of temperature at fixed external field in the presence of skyrmions and 2) the skyrmion fluctuations of energy around the average energy $<E>$.

The evaluation of the mean square fluctuations of the skyrmion energy is based on the calculation of the partition function yielding $<\delta E^2> \simeq k_B T^3 \left(2\frac{\partial S}{\partial T} + T\frac{\partial^2 S}{\partial T^2}\right)$ with $<\delta E^2> = <E^2> - <E>^2$ and $<E>$ being the average energy. The key result is that skyrmion mean fluctuations of energy have an entropic dependence. By using Eq. (6), we get $<\delta E^2> \simeq \frac{1}{2}(k_B T)^2$, namely a proportionality to the square of the temperature (see the Appendix B for the details). This result is general and does not depend on the external bias field. The factor ½ is consistent with the fact that, unlike the case of the ideal gas where particle velocity is a vector with three degrees of freedom, here there is one degree of freedom represented by the skyrmion.

In Fig. 6, we display the mean square fluctuations of the energy of the skyrmion diameter distribution as a function of temperature. One notes that there is a quadratic increase of $<\delta E^2>$ with temperature, thus showing a behavior similar to that of the mean square fluctuations of energy of gas particles in a canonical ensemble.

To establish whether the skyrmion fluctuations of energy are relevant, we compare their order of magnitude with that of the average energy calculating the fractional mean square fluctuation of the energy: $<\delta E^2>/<E>^2 \simeq \frac{1}{2}(k_B T)^2/<E>^2$. For the system studied and the range of temperatures considered, looking at Fig. 5 on average $<\delta E^2> \approx 10^{-41} J^2$. As $<E>^2 \approx 10^{-34} J^2$ we get $\frac{<\delta E^2>}{<E>^2} \simeq 10^{-7}$ resulting in fluctuations of energy that are about three orders of magnitude smaller than the average energy, $\sqrt{<\delta E^2>} \ll <E>$. In principle, the skyrmion diameters distribution is supposed in contact with a heat reservoir forming a canonical ensemble and the average energy is determined by the temperature of the heat reservoir itself. In a canonical ensemble, the total energy is not conserved and, therefore, for this special case, the skyrmion diameter population exhibits fluctuations of energy. However, due to the high number of degrees of freedom represented by the skyrmions population with different diameters, we can suppose that the fluctuations of energy are very small treating, in a first instance, the ensemble in the same way as a microcanonical ensemble. This hypothesis is confirmed by the calculation of the energy fluctuations that are very small if compared to the average energy as shown above.

Analogously to what occurs to the number of non-interacting molecules of an ideal gas, that varies continuously as a function of velocity and possess only translational kinetic energy, also the skyrmion diameters, during thermal annealing, fluctuate independently along the radial direction, leading to continuous and infinitesimal changes of skyrmion size. Another common property is represented by confinement. Like gas molecules obeying MB statistics are confined in a box, magnetic skyrmions of different diameters are formed in a confined magnetic system.

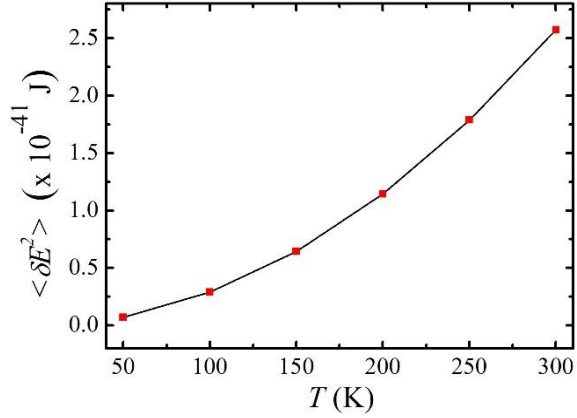

Fig. 6: Mean square fluctuations of energy of the average skyrmion energy as a function of temperature.

## IV.    CONCLUSIONS

In summary, we have shown that it is possible to describe the statistical behavior of the skyrmion diameter in presence of thermal fluctuations by using a statistical thermodynamic analogy with the non-interacting molecules of an ideal diluted gas. This analogy is valid in the region of energetic metastability for the skyrmion where the following hypotheses are verified: (i) the skyrmion energy close to the minimum exhibits a parabolic profile and (ii) the skyrmion diameter population follows a MB statistical distribution. We confirmed those hypotheses with the results of micromagnetic simulations. Therefore, we have developed an analytical model able to describe the statistical behavior of the skyrmion diameter (average value, standard deviation, and distribution) as well as to calculate the configurational entropy of a skyrmion linked to the thermal breathing mode and internal deformations. In the low-temperature limit, the average skyrmion diameter has a linear dependence on $T$, while the configurational entropy shows a deviation from the linear behavior. From the calculation of the partition function, we have expressed fluctuations of energy as a function of entropy only. Those results can be used to study phase transitions involving skyrmions and their relaxation properties.

V.   ACKNOWLEDGEMENTS

R.T. and G.F. thank the project "ThunderSKY" funded from the Hellenic Foundation for Research and Innovation and the General Secretariat for Research and Technology, under Grant No. 871. R.Z. acknowledges support by Gruppo Nazionale per la Fisica Matematica (GNFM) and Istituto Nazionale di Alta Matematica (INdAM) "F. Severi." O.C.-F. also acknowledges the Spanish Ministry of Economy and Competitiveness under the project FIS2016–78591-C3-3-R and the support from the Messina University with its Visiting Research program. G. F., O.C.-F., and M.C. would like to acknowledge networking support by the COST Action CA17123 "Ultrafast opto magneto electronics for non-dissipative information technology". R.Z. thanks Daniele Pinna for fruitful discussions.

APPENDIX A: CALCULATION OF THE CONFIGURATIONAL ENTROPY

We recall Eq. (5) of the main text:

$$H_0 = \frac{\pi}{2} \int_0^\infty dD_{\text{sky}} D_{\text{sky}}^2 \, f_0 \ln f_0. \tag{A1}$$

The computation of the integral yields:

$$H_0 \simeq \frac{C_{\text{sky}}}{4}\left(\frac{\pi}{\beta a}\right)^{\frac{3}{2}}\left[\left(1+2\beta a <D_{\text{sky}}>^2\right)\ln\left(C_{\text{sky}} <V>\right) - \frac{1}{4}\left(3+2\beta a <D_{\text{sky}}>^2\right)\right], \tag{A2}$$

where $<V> = \frac{1}{4}\pi <D_{\text{sky}}>^2 \, t$ is the average skyrmion volume. We get the constant $C_{\text{sky}}$ via a normalization condition $C_{\text{sky}} = \frac{2}{\pi}\left(\int_0^\infty dD_{\text{sky}} D_{\text{sky}}^2 \, e^{-\beta\left[a\left(D_{\text{sky}} - <D_{\text{sky}}>\right)^2\right]}\right)^{-1}$ resulting in

$C_{\text{sky}} \simeq 4\left(\dfrac{a}{\pi k_B T}\right)^{\frac{3}{2}} \dfrac{k_B T}{k_B T + 2a <D_{\text{sky}}>^2}$. We substitute $C_{\text{sky}}$ into $H_0$, via $\beta = \dfrac{1}{k_B T}$, getting

$$H_0 \simeq \frac{C_{\text{sky}}}{4}\left(\frac{\pi k_B T}{a}\right)^{\frac{3}{2}}\left[\left(\frac{k_B T + 2a <D_{\text{sky}}>^2}{k_B T}\right)\ln\left(C_{\text{sky}} <V>\right) - \frac{1}{4}\left(\frac{3k_B T + 2a <D_{\text{sky}}>^2}{k_B T}\right)\right]. \tag{A3}$$

The skyrmion entropy at the thermodynamic equilibrium is calculated as $S = -k_B H_0$ after substituting $C$ and $<V>$:

$$S \simeq k_B \left[ \ln\left( \frac{(k_B T)^{\frac{3}{2}} + 2(k_B T)^{\frac{1}{2}} a <D_{sky}>^2}{a^{\frac{3}{2}} <D_{sky}>^2 t} \right) + \frac{1}{4}\left( \frac{3k_B T + 2a <D_{sky}>^2}{k_B T + 2a <D_{sky}>^2} \right) \right] + S_0, \quad (A4)$$

with $S_0 = \frac{1}{2} k_B \ln \pi$. Eq. (A4) is Eq. (6) of the main text.

## APPENDIX B: CALCULATION OF THE MAGNETIZATION DISTRIBUTION AND OF THE ENERGY FLUCTUATIONS

The skyrmion entropy is crucial to determine: 1) the magnetization distribution as a function of temperature at fixed external bias field and 2) the skyrmion fluctuations of energy around the average energy $<E>$.

The calculation of the magnetization distribution is based on the hyperbolic law expressing the average skyrmion diameter as a function of $B = \mu_0 H_{ext}$ at fixed $T$ as observed by Romming for $T = 4.2$ K *et al.* [57] and confirmed by our model at larger temperatures:

$$<D_{sky}(B)> = \frac{B_0}{B+B_0} <D_{sky\ B=0}>, \quad (B1)$$

with $B$ expressed here in $J$ ($B \rightarrow \mu_0 H_{ext} M_s V$) and $B_0$ a parameter also expressed in $J$ that can be obtained from a fit to experimental data [57].

First, we calculate the entropy dependence on the bias external field at fixed temperature. Substituting Eq.(B1) into Eq.(6) we get the skyrmion entropy dependence on $B$ at fixed $T$:

$$S(B) \simeq k_B \left[ \ln\left[ \frac{(k_B T)^{\frac{3}{2}} + 2(k_B T)^{\frac{1}{2}} a\left(\frac{B_0}{B+B_0}<D_{sky\ B=0}>\right)^2}{(a)^{\frac{3}{2}}\left(\frac{B_0}{B+B_0}<D_{sky\ B=0}>\right)^2 t} \right] + \frac{1}{4}\left( \frac{3k_B T + 2a\left(\frac{B_0}{B+B_0}<D_{sky\ B=0}>\right)^2}{k_B T + 2a\left(\frac{B_0}{B+B_0}<D_{sky\ B=0}>\right)^2} \right) \right] + S_0. \quad (B2)$$

We immediately find, using Maxwell's relation $\left(\frac{\partial S}{\partial B}\right)_T = \left(\frac{\partial M}{\partial T}\right)_B$:

$$\left(\frac{\partial M}{\partial T}\right)_B = 2M_s \left[ k_B^2 T^2 \left( \frac{(B+B_0)\left(k_B T (B+B_0)^2 + 3a <D_{sky\ B=0}>^2 B_0^2\right)}{\left(k_B T (B+B_0)^2 + 2a <D_{sky\ B=0}>^2 B_0^2\right)^2} \right) \right], \quad (B3)$$

with $a = a(T)$. For $B = 0$ J ($\mu_0 H_{ext}= 0$ mT), the magnetization distribution as a function of temperature at different $T$ is:

$$\left(\frac{\partial M}{\partial T}\right)_B = 2\frac{M_s}{B_0}\left[k_B^2 T\left(\frac{k_B T + 3a <D_{sky\,B=0}>^2}{\left(k_B T + 2a <D_{sky\,B=0}>^2\right)^2}\right)\right]. \tag{B4}$$

We immediately get the asymptotic behavior of the magnetization distribution for very high temperatures, $\lim_{T\to\infty}\left[\left(\frac{\partial M}{\partial T}\right)_{B=0}\right] = 2k_B \frac{M_s}{B_0}$.

The integration over $T$ in the interval $[0, T_R]$ with $T_R$ the reference temperature gives the magnetization at the temperature $T_R$ in the absence of an external bias field:

$$M(T_R)_{B=0} = 2\frac{M_s}{B_0}\left[k_B T_R \frac{k_B T_R + a <D_{sky\,B=0}>^2}{k_B T_R + 2a <D_{sky\,B=0}>^2} - a<D_{B=0}>^2 \ln\left(\frac{k_B T_R + 2a <D_{sky\,B=0}>^2}{2a <D_{sky\,B=0}>^2}\right)\right], \tag{B5}$$

The magnetization depends on two contributions and both of them are functions of the relevant parameters characterizing skyrmion energy, $a$ and $<D_{B=0}>$. We now outline the calculation of 2), the skyrmion fluctuations of energy. Let us start from the thermodynamic relation linking the average energy with the partition function $Z$, $<E> = -\frac{\partial \ln Z}{\partial \beta}$. This implies that $<E>^2 = \left(\frac{\partial \ln Z}{\partial \beta}\right)^2$. We now express the average of the square of the energy as a function of the partition function. We write the average of the square of the energy in the continuous limit as:

$$<E^2> = \frac{C}{Z}\int [E(x)]^2 e^{-\beta E(x)} dx, \tag{B6}$$

where $C$ is a normalization constant and $x$ is a generic variable. We cast Eq. (S19) in the form $<E^2> = \frac{C}{Z}\int \frac{\partial^2}{\partial \beta^2}\left(e^{-\beta E(x)}\right)dx = \frac{1}{Z}\frac{\partial^2 Z}{\partial \beta^2} = \frac{\partial^2 \ln Z}{\partial \beta^2} + \left[\frac{1}{Z}\left(\frac{\partial Z}{\partial \beta}\right)\right]^2 = \frac{\partial^2 \ln Z}{\partial \beta^2} + \left(\frac{\partial \ln Z}{\partial \beta}\right)^2$. Hence, we get $<E^2> - <E>^2 = \frac{\partial^2 \ln Z}{\partial \beta^2} + \left(\frac{\partial \ln Z}{\partial \beta}\right)^2 - \left(\frac{\partial \ln Z}{\partial \beta}\right)^2 = \frac{\partial^2 \ln Z}{\partial \beta^2}$. The square fluctuations of energy take the general form:

$$<\delta E^2> = \frac{\partial^2 \ln Z}{\partial \beta^2}, \tag{B7}$$

being $<\delta E^2> = <E^2> - <E>^2$.

To get an explicit expression of the fluctuations of energy, we write down the partition function of the system that can be regarded as a controlling function able to determine the average energy of the system itself. In the continuum limit, we write the partition function as an integral over Boltzmann factors, $Z = C_N \int e^{-\beta E(x)} dx$ where $E(x)$ is the energy depending on the general continuous variable $x$ and $C_N$ is a normalization constant. As stated above, in principle the system under study represented by the ensemble of skyrmion diameters is a canonical ensemble with a very high number of degrees of freedom whose fluctuations of energy are thus very small. This allows assuming that $E(x)$ is very close to the average energy $<E>$. This corresponds to treat the partition function of a canonical ensemble as that of a microcanonical one. Hence, without loss of generality, we express $Z \simeq W e^{-\beta <E>}$ where $W$ is the statistical multiplicity of the energy level having value $<E>$ and $<E> = -\frac{\partial \ln Z}{\partial \beta}$. This allows for writing, via $S = k_B \ln W$, the simple relation:

$$\ln Z = \frac{S}{k_B} - \frac{<E>}{k_B T}. \tag{B8}$$

We now derive the mean square fluctuations of energy $<\delta E^2>$ around the mean value $<E>$ that, for a microcanonical ensemble, are very small if compared to the average energy. Taking into account Eq. (S15) and Eq. (S16) and expressing the mean square fluctuations of energy as a function of $T$ yields:

$$<\delta E^2> = k_B T^3 \left( 2\frac{\partial S}{\partial T} + T \frac{\partial^2 S}{\partial T^2} \right) - k_B^3 T^5 \frac{\partial^2 <E>}{\partial T^2} - 2 k_B^3 T^4 \frac{\partial <E>}{\partial T} - 2 k_B^3 T^3 <E>. \tag{B9}$$

Simple considerations allow us to simplify Eq. (S17). Indeed, the average energy is linear in $T$ so that its second derivative equals zero. Furthermore, the three terms proportional to $k_B^3$ proportional to the average energy and its first and second derivatives with respect to temperature are much smaller than the first term and can be safely neglected. Hence:

$$<\delta E^2> \simeq k_B T^3 \left( 2\frac{\partial S}{\partial T} + T \frac{\partial^2 S}{\partial T^2} \right). \tag{B10}$$

Mean fluctuations of energy around the mean energy of the skyrmion diameters population depend on entropy. By replacing the entropy expressed in Eq. (S8), we get:

$$<\delta E^2> = \frac{1}{2} k_B^2 T^2 + 2\frac{k_B^3 T^3}{2a<D_{sky}>^2 + k_B T} - \frac{k_B^4 T^4}{\left(2a<D_{sky}>^2 + k_B T\right)^2} + 2\frac{a<D_{sky}>^2 k_B^3 T^3}{\left(2a<D_{sky}>^2 + k_B T\right)^2} - 2\frac{a<D_{sky}>^2 k_B^4 T^4}{\left(2a<D_{sky}>^2 + k_B T\right)^3}. \tag{B11}$$

The first term on the second member is much larger than the other terms so that:

$$<\delta E^2> \simeq \frac{1}{2} k_B^2 T^2. \tag{B12}$$

# SUPPLEMENTARY INFORMATION

# Configurational entropy of magnetic skyrmions as an ideal gas

R. Zivieri[1*], R. Tomasello[2#], O. Chubykalo-Fesenko[3], V. Tiberkevich[4], M. Carpentieri[5], and G. Finocchio[1§]

[1]*Department of Mathematical and Computer Sciences, Physical Sciences and Earth Sciences, University of Messina, 98166 Messina, Italy*

[2]*Institute of Applied and Computational Mathematics, Foundation for Research and Technology, GR 700 13 Heraklion, Crete, Greece*

[3]*Instituto de Ciencia de Materiales de Madrid, CSIC, Cantoblanco, Madrid, Spain*

[4]*Department of Physics, Oakland University, Rochester, Michigan 48309, USA*

[5]*Department of Electrical and Information Engineering, Politecnico di Bari, Bari, Italy*

*zivieri@fe.infn.it

#rtomasello@iacm.forth.gr

§gfinocchio@unime.it

## Supplementary note 1

The micromagnetic computations are performed by a state-of-the-art micromagnetic solver which numerically integrates the Landau-Lifshitz-Gilbert (LLG) equation by applying the time solver scheme Adams-Bashforth [1]:

$$(1+\alpha_G^2)\frac{d\mathbf{m}}{d\tau} = -(\mathbf{m}\times\mathbf{h}_{\text{eff}}) + \alpha_G \mathbf{m}\times(\mathbf{m}\times\mathbf{h}_{\text{eff}}) \quad \text{(S1)}$$

where $\alpha_G$ is the Gilbert damping, $\mathbf{m} = \mathbf{M}/M_s$ is the normalized (reduced) magnetization, and $\tau = \gamma_0 M_s t$ is the dimensionless time, with $\gamma_0$ being the gyromagnetic ratio, and $M_s$ the saturation magnetization. $\mathbf{h}_{\text{eff}}$ is the normalized effective magnetic field, which includes the exchange, magnetostatic, anisotropy and external fields, as well as the interfacial DMI and the thermal field. The interfacial DMI contribution $\mathbf{h}_{\text{DMI}}$ is obtained from the functional derivative of the DMI energy density $\varepsilon_{\text{DMI}} = D\left[m_z \nabla\cdot\mathbf{m} - (\mathbf{m}\cdot\nabla)m_z\right]$ under the hypothesis of thin film $\left(\frac{\partial \mathbf{m}}{\partial z} = 0\right)$ [2,3] as

$$\mathbf{h}_{\text{DMI}} = -\frac{2D}{\mu_0 M_S}\left[(\nabla\cdot\mathbf{m})\hat{z} - \nabla m_z\right], \quad \text{(S2)}$$

with $D$ being the parameter taking into account the intensity of the DMI, $m_z$ the out-of-plane component of the normalized magnetization, $\mu_0$ the vacuum permeability, and $\hat{z}$ the unit vector along the out-of-plane direction. The DMI affects the boundary conditions of the ferromagnetic sample in the following way $\frac{d\mathbf{m}}{dn} = \frac{D}{2A}(\hat{z}\times\mathbf{n})\times\mathbf{m}$, where $\mathbf{n}$ is the unit vector normal to the edge, and $A$ is the exchange constant.

## Supplementary note 2

In this section, we derive the most probable value of skyrmion diameter population at a given temperature $D_{sky}$ taking into account the analogy with the statistical behavior of an ideal gas that was discussed in the main text. For this latter, it is customary to define the most probable speed of a molecule by the value of speed v at which the Maxwell-Boltzmann (MB) distribution attains a maximum. According to our model the skyrmion energy is fitted via a parabola $E \simeq a(D_{sky} - D_{0\,sky})^2 + b$ with $a = \frac{1}{2}\,d^2E/dD_{sky}^2$ proportional to the energy curvature, $D_{0\,sky}$ the equilibrium diameter at a given temperature, $D_{0\,sky} = D_{0\,sky}(T)$, and $b = E_{sky}(D_{sky} = D_{0\,sky})$ corresponding to the energy minimum $E_{sky}^{min}$ at a given temperature. In our model, both $a$ and $b$ are temperature-dependent coefficients $a(T)$ and $b(T)$.

Eq. (1) of the main text expressing the skyrmion diameter distribution reads:

$$\frac{dn}{dD_{sky}} = C_{sky}\, D_{sky}^2\, e^{-\beta a(D_{sky} - D_{0\,sky})^2}, \tag{S3}$$

where $C$ is the normalization constant, $\beta = \dfrac{1}{k_B T}$ with $k_B = 1.38 \times 10^{-23}$ J/K the Boltzmann constant and $T$ the temperature.

From the minimization of Eq. (S3), one gets the displacement law of the diameter with the temperature, namely:

$$D_{sky}^{mp}(T) = \frac{1}{2}\left(D_{0\,sky} + \sqrt{D_{0\,sky}^2 + \frac{4k_B T}{a}}\right), \tag{S4}$$

where the superscript "mp" stands for "most probable" corresponding to the maximum of the distribution and $D_{sky}^{mp}(T)$ is the most probable value of the skyrmion diameter. This law expresses the displacement of the maximum of the distribution as a function of $T$. It can be noted that the diameter $D_{sky}^{mp}(T)$ depends on the equilibrium diameter $D_{0\,sky}(T)$ at a given $T$ and that, at low temperatures expanding to the first order, $D_{sky}^{mp}(T)$ has a linear dependence on $T$ assuming $D_{0\,sky}(T)$ $D_{0\,sky}(T) = D_{0\,sky}(T = 0\,\text{K}) + d\,T$ (both via an analytical and a micromagnetic check) with $d$ a coefficient expressed in m/K.

Straightforwardly, setting $D_{sky}^{mp}(T) = v_{mp}(T)$ ($v_{mp}(T)$ is the most probable speed), $D_{0\,sky} = 0$ and $a = \frac{1}{2} m$ (*m* the mass of a gas molecule), we get the well-known result $v_{mp}(T) = \sqrt{2k_B T/m}$.

Note that $1 < \frac{<D_{sky}(T)>}{D_{sky}^{mp}(T)} < 1.1$ with $<D_{sky}(T)>$ a ratio depending weakly on temperature with $\frac{<D_{sky}(T)>}{D_{sky}^{mp}(T)} \to 1$ as $T \to 0$. This result is slightly different from that of gases where the ratio $\frac{<v(T)>}{v_{mp}(T)} = 1.13$ for each temperature with $<v(T)>$ the average velocity.